\RequirePackage{fix-cm}
\documentclass[natbib]{svjour3}
\journalname{Celestial Mechanics and Dynamical Astronomy }
\smartqed  
\usepackage{graphicx}
\begin{document}

\title{Motion of particles on a $z=2$ Lifshitz black hole background  in 3+1 dimensions}

\author{ Marco Olivares   \and
         Yerko V\'asquez  \and \\
         J. R. Villanueva \and
         Felipe Moncada
}
\institute{ Marco Olivares \at
           Instituto de F\'{\i}sica, Pontificia Universidad
Cat\'{o}lica de Valpara\'{\i}so, Av. Universidad 330, Curauma,
Valpara\'{\i}so, Chile.\\
               \email{marco.olivaresrubilar@gmail.com}
          \and
           Yerko V\'asquez \at
              Departamento de F\'isica, Facultad de Ciencias, Universidad de La Serena, 
              Av. Cisternas 1200, La Serena, Chile.\at
	      Departamento de Ciencias F\'isicas, Facultad de Ingenier\'ia y
	      Ciencias, Universidad de La Frontera, Av. Francisco
	      Salazar 01145, Casilla 54-D, Temuco, Chile. \\
              \email{yerko.vasquez@ufrontera.cl}
           \and
          J. R. Villanueva \at
          Instituto de F\'{\i}sica y Astronom\'{\i}a, Universidad de Valpara\'{\i}so, 
          Gran Breta\~na 1111, Playa Ancha, Valpara\'{\i}so, Chile.\at
          Centro de Astrof\'{\i}sica de Valpara\'{\i}so,
Gran Breta\~na 1111, Playa Ancha, Valpara\'{\i}so, Chile.\\
\email{jose.villanuevalob@uv.cl}
 \and
              Felipe Moncada \at
              Departamento de Ciencias F\'isicas, Facultad de Ingenier\'ia y
Ciencias, Universidad de La Frontera, Av. Francisco
Salazar 01145, Casilla 54-D, Temuco, Chile. \\
              \email{f.moncada02@ufromail.cl}
}

\date{Received: date / Accepted: date}

\maketitle

\begin{abstract}
The object of study is the geodesic structure of a $z=2$ Lifshitz black hole in 3+1
space-time dimensions, which is an exact solution to the
Einstein--scalar--Maxwell theory. The motion of massless and massive particles in this background is researched using the standard Lagrangian procedure. Analytical expressions are obtained for radial and angular motions of the test particles, where the polar trajectories are given in terms of the $\wp$--Weierstra{\ss} elliptic function. 
It will be demonstrated that an external observer can see that photons with radial motion arrive at spatial infinity in a finite coordinate time. For particles with non-vanished angular momentum, the motion is studied on the invariant plane $\phi= \pi/2$ and, it is shown that bounded orbits are not allowed for this space-time on this plane. These results are consistent with other recent studies on Lifshitz black holes.
\end{abstract}

\section{Introduction}
\label{intro}
Motivated by generalizations to other areas of physics of the AdS/CFT correspondence \citep{Maldacena:1997re} , i.e., the duality between the geometry of a $d$-dimensional anti-de Sitter (AdS) space and a $d-1$-dimensional conformal field theory (CFT), Lifshitz space-times
have received considerable attention of late. They represent
gravity duals of non-relativistic systems that appear in condensed matter
physics \citep{Kachru:2008yh,Hartnoll:2009ns} with an anisotropic
scaling symmetry $t\rightarrow \lambda ^{z}t$, $x\rightarrow \lambda x$,
where $z$ is the dynamical exponent accounting for the different scale
transformation between the temporal and spatial coordinates. These
space-times are described in 3+1 dimensions by the metrics

\begin{equation}
ds^{2}=\ell^{2}\left( -r^{2z}dt^{2}+\frac{dr^{2}}{r^{2}}+r^{2}d\vec{x}%
^{2}\right) ,  \label{ecuacion}
\end{equation}%
where $\vec{x}$ represents a two-dimensional vector and the radial coordinate $%
r$ scales according to $r\rightarrow r/\lambda $. It is worth mentioning
that for $z=1$ the above metric reduces to the usual four-dimensional AdS
metric. Black holes with asymptotic behavior given by these anisotropic scale invariant
space-times represent gravity duals of condensed matter systems at finite
temperature. Asymptotically Lifshitz black hole solutions have been
reported in \citet{Balasubramanian:2009rx}, \citet{AyonBeato:2009nh}, \citet{mann} and \citet{Dehghani:2010kd}.
Some thermodynamic aspects of these black holes have also been 
studied in \citet{Devecioglu:2011yi}, \citet{Myung:2012cb}, and
\citet{oai:arXiv.org:1203.1367}.

This paper focuses on the geodesic structure of a 3+1 dimensional
black hole presented by \citet{Taylor:2008tg}, the asymptotic behavior of which is
given by Eq.(\ref{ecuacion}) with dynamical exponent $z=2$, which emerges
as an exact solution of the Einstein-scalar-Maxwell theory \citep{Taylor:2008tg}, \citep{Pang:2009wa}. Geodesic studies of  the $z=2$ topological Lifshitz black
hole in 3+1 dimensions and of the $z=3$ Lifshitz black hole in 2+1
dimensions, which has been found to be a solution to the New Massive Gravity theory, have been
reported recently in \citet{Olivares:2013uha} and \citet{Cruz:2013ufa},
respectively. In this investigation  the motion of massless and
massive particles is studied in a similar black hole background using the standard
Lagrangian procedure \citep{Olivares:2013uha,Cruz:2013ufa,Olivares,villanueva1,ov13}. 
The effective potential analysis provides the means to describe the motion of particles along null and time-like geodesics. The exact solutions of the geodesic equations are presented. The analytical expressions for the radial geodesics are given as expressions of the proper and coordinate times. Moreover, the equations for the angular motion of the test particle are provided through the $\wp$--Weierstra{\ss} %
  function.
 
The paper is organized as follows: In section II the geodesic equations are obtained for (massless and massive) particles in the space-time found in the 3+1 dimensional space-time found in \citet{Taylor:2008tg} and \citet{Pang:2009wa}.
Then their radial and angular motions are examined. Finally, in section III the results and conclusions are discussed.

\section{Geodesic structure}
Let us consider the action of the
Einstein-scalar-Maxwell theory, which is given by \citep{Taylor:2008tg,Pang:2009wa}

\begin{equation}
\mathcal{I}_{EsM}=\frac{1}{16\pi G}\int d^{4}x\sqrt{-g}\left( R-2\Lambda -\frac{1}{2}%
\partial _{\mu }\varphi\, \partial ^{\mu }\varphi -\frac{1}{4}e^{\tilde{\lambda}\varphi
}F_{\mu \nu }F^{\mu \nu }\right),
\label{g1}
\end{equation}%
where $\Lambda $ is the cosmological constant, $\varphi $ is a massless scalar
field, and $F_{\mu \nu }$ corresponds to the Maxwell field.
An analytical $z=2$ asymptotically Lifshitz black hole solution to this
theory with a flat transverse section is given by the metric 
\citep{Amado:2011nd,Brynjolfsson:2009ct}

\begin{equation}
ds^{2}=\ell^{2}\left[ -r^{4}\left( 1-\frac{r_{+}^{4}}{r^{4}}\right) dt^{2}+%
\frac{dr^{2}}{r^{2}\left( 1-\frac{r_{+}^{4}}{r^{4}}\right) }+r^{2}\left(
d\theta ^{2}+\theta ^{2}d\phi ^{2}\right) \right],
\label{g2}
\end{equation}%
where the curvature radius of the Lifshitz black hole, $\ell$, is related
to the cosmological constant $\Lambda=-6/\ell^2$.
The event horizon is located at
\begin{equation}
r_{+}=\left(\frac{8\,\pi\,G\,\mathcal{M}}{\ell^2V_2}\right)^{1/4},
\label{g2.1}
\end{equation}
where $\mathcal{M}$ is the mass of the corresponding black hole determined by
 the Euclidean action approach, and
$V_{2}$ is the volume of two-dimensional spatial directions.
While, the
scalar and Maxwell fields are given by the expressions

\begin{equation}\label{g3}
e^{\tilde{\lambda}\varphi }=\frac{1}{r^{4}}\,(\tilde{\lambda}^{2}=4),%
\quad F_{tr}=2\sqrt{2}\ell\,r^{3}.
\end{equation}%
\citet{Myung:2012cb} determined the thermodynamical properties of this Lifshitz 
black hole and presented a stability analysis
 considering the scalar field perturbation of this black 
 hole. In particular, it was found that the 
 temperature $T_H$, the Bekenstein--Hawking 
 entropy $S_{BH}$, the heat capacity $C$,  and 
the Helmholtz free energy $F$ are given by
\begin{equation}\label{g4}
T_{H}=\frac{r_{+}^{2}}{\pi },\quad S_{BH}=\frac{\ell^{2}V_{2}}{4G}r_{+}^{2},\quad C=\frac{%
2\ell^{2}V_{2}r_{+}^{2}}{8G},\quad F=-\frac{2\ell^{2}V_{2}r_{+}^{4}}{16\pi G}.
\end{equation}

The curvature scalar, the principal quadratic invariant of the Ricci tensor
and the Kretschmann scalar of this space-time are given by the following
expressions

\begin{eqnarray}\label{g5}
R&=&\frac{22}{\ell^{2}}+\frac{2r_{+}^{4}}{\ell^{2}r^{4}}, \\ \label{g6}
R_{\mu \nu }R^{\mu \nu }&=&\frac{4\left(
33r^{8}+r_{+}^{8}+6r_{+}^{4}r^{4}\right) }{\ell^{4}r^{8}}, \\ \label{g7}
R_{\mu \nu \rho \sigma }R^{\mu \nu \rho \sigma }&=&\frac{4\left(
3\ell^{2}r_{+}^{8}+2r^{2}+6\ell^{2}r_{+}^{4}r^{4}+23\ell^{2}r^{8}\right) }{\ell^{6}r^{8}}%
.
\end{eqnarray}
These curvature invariants are regular on the 
event horizon, $r_{+}$, and therefore this surface
only expresses a singularity of the 
coordinates used to define the metric (\ref{g2}).
However, all these scalars diverge in $r = 0$, 
and thus this point corresponds 
to a curvature singularity.

In order to study the motion of test particles in the background (\ref{g2}),
the standard Lagrangian approach is used (\citet{Olivares,villanueva1,ov13}).
The corresponding Lagrangian is

\begin{equation}
2\mathcal{L} =\ell^{2}\left(-\left( r^{4}-r_{+}^{4}\right) \dot{t}^{2}+%
\frac{r^{2}\dot{r}^{2}}{\left( r^{4}-r_{+}^{4}\right) }%
+r^{2}\left( \dot{\theta}^{2}+\theta ^{2}\dot{\phi}^{2}\right)\right) =-m\,\ell^{2}.
\label{g6}
\end{equation}%
Here, the dot refers to a derivative with respect to an affine parameter, $\tau $,
along the trajectory, and, by normalization, $m=0\,(1)$ for massless (massive)
particles.
Since $(t, \phi)$ are cyclic coordinates, their corresponding
conjugate momenta $(\Pi _{t}, \Pi _{\phi })$ satisfy the following 
relations:
\begin{eqnarray}\label{g7}
\frac{d}{d\tau}\frac{\partial \mathcal{L}}{\partial \dot{t}}&=&\frac{d\,\Pi _{t}}{d\tau}=
\frac{d[-\ell^{2}\left( r^{4}-r_{+}^{4}\right) \dot{t}\,]}{d\tau}=0
,\\ \label{g7.1}
\frac{d}{d\tau}\frac{\partial \mathcal{L}}{\partial \dot{\phi}}&=&\frac{d\,\Pi_{\phi }}{d\tau} =\frac{d[\ell^{2}r^{2}\theta ^{2}\dot{\phi}]}{d\tau}=0.
\end{eqnarray}%
Furthermore, the equation of motion
associated with the $\theta$ coordinate becomes
\begin{equation}\label{g7.2}
\frac{d}{d\tau}\frac{\partial \mathcal{L}}{\partial \dot{\theta}}=\frac{d\,\Pi_{\theta }}{d\tau} =\frac{d\,(\ell^2\,r^2\,\dot{\theta})}{d\tau}=\ell^2\,r^2\,\theta\,\dot{\phi}^2.
\end{equation}
Conversely,  the flat metric (\ref{g2}) 
admits the following Killing vectors field: 
\begin{center}
\begin{itemize}
\item the \textit{time-like Killing vector} $\xi=\partial_{t}$ 
is related to the \textit{stationarity} of the metric. The conserved 
quantity is given by
\begin{equation}\label{eqE}
g_{\alpha \beta }\,\xi^{\alpha}\,u^{\beta}
=-\ell^{2}\,(r^4-r_{+}^{4})\,\dot{t}=-\ell^{2}\,\sqrt{E},
\end{equation}
 where $E$ 
is a constant of motion that cannot be associated with the total energy of the test particles
because this space-time is asymptotically Lifshitz, and not
flat.
\item the \textit{space-like Killing vectors}
$\chi_0=\partial_{\phi}$, $\chi_1= \theta^{-1}\sin \phi\,
\partial_{\phi}-\cos \phi\,\partial_{\theta}$, and
$\chi_2=\theta^{-1}\cos \phi\,
\partial_{\phi}+\sin \phi\,\partial_{\theta}$, which are related to the
axial symmetry of the metric. The conserved quantities
are given by
\begin{eqnarray}\label{cca0}
g_{\alpha \beta }\,\chi_{0}^{\alpha}\,u^{\beta}
&=&\ell^{2}\,r^{2}\,\theta^2\,\dot{\phi}
=C_0, \\ \label{cca1}
g_{\alpha \beta }\,\chi_{1}^{\alpha}\,u^{\beta}
&=&\ell^{2}\,r^{2}\,(\sin\phi\,\dot{\theta}+\theta\, \cos\phi\,\dot{\phi})
=C_1, \\ \label{cca2}
g_{\alpha \beta }\,\chi_{2}^{\alpha}\,u^{\beta}
&=&\ell^{2}\,r^{2}\,(-\cos\phi\,\dot{\theta}+\theta\, \sin\phi\,\dot{\phi})
=C_2,
\end{eqnarray}
where $C_0$, $C_1$ and $C_2$ are constants associated with the angular momentum
of the particles.
\end{itemize}
\end{center}
This last point implies that the motion can be restricted on an invariant plane, i.e., $\dot{\phi}=0 \rightarrow \phi= const.$ , which, for simplicity we set at $\phi= \pi/2$.
Therefore, $C_0=C_2=0$, so Eqs. (\ref{eqE}) and (\ref{g7.2}, \ref{cca1}) lead
to the following expressions

\begin{equation}\label{g8}
\dot{t}=\frac{\sqrt{E}}{\left( r^{4}-r_{+}^{4}\right) },\qquad
\,\dot{\theta}=\frac{L}{r^{2}},
\end{equation}
where $L=C_1/\ell^2$, which denotes the angular momentum about 
an axis normal to the invariant plane.

These relations together with Eq. (\ref{g6}) make it possible  to obtain the following
differential equations:

\begin{eqnarray}\label{g9}
\left(\frac{dr}{d\tau}\right)^{2}&=&\frac{1}{r^{2}}\left( E-V_{eff}\left( r\right) \right),\\ \label{g10}
\left(\frac{dr}{dt}\right)^{2}&=&\frac{(r^4-r_+^4)^2}{r^{2}}\,\left( \frac{E-V_{eff}\left( r\right)}{E} \right),\\ \label{g11}
\left(\frac{dr}{d\theta}\right)^{2}&=&\frac{r^{2}}{L^2}\left( E-V_{eff}\left( r\right) \right),
\end{eqnarray}%
where the effective potential, $V_{eff}\left( r\right) $, reads

\begin{equation}
V_{eff}\left( r\right) =\left( r^{4}-r_{+}^{4}\right)
\left( m+\frac{L^{2}}{r^{2}}\right) .  \label{g12}
\end{equation}%
In the following sections, based on this effective potential, the
geodesic structure of the space-time characterized by the metric  (\ref{g2})
is analyzed.

\subsection{Null geodesics}
\begin{figure}[!h]
 \begin{center}
  \includegraphics[width=85mm]{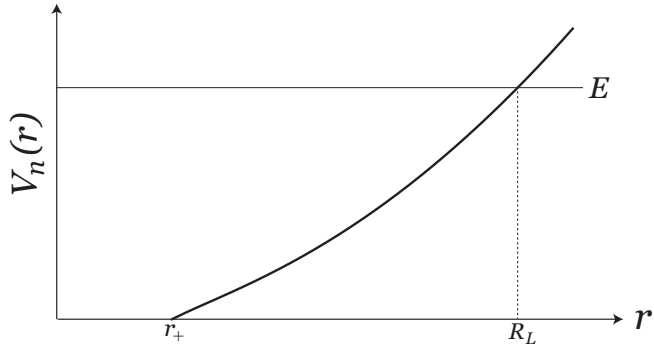}
 \end{center}
 \caption{Effective potential for massless particles.
 The plot shows that photons with 
 $L\neq 0$ cannot escape to the spatial infinity.
 The value of the constant of motion $E$ determines
 the position of the turning point $R_L$.}
 \label{f1}
\end{figure}

 In order to study the motion of massless particles,
let us consider the effective potential (\ref{g12}) with
$m=0$, so it can be expressed by 
\begin{equation}\label{n1}
  V_{n}\left( r\right) =L^{2}\left( \frac{r^{4}-r_{+}^{4}}{r^{2}}\right).
\end{equation}
A typical graph of this effective 
potential is shown in Fig.  \ref{f1}, for the arbitrary
value of $L\neq0$. From this plot
it is easy to see that photons cannot escape 
to spatial infinity, but that the confined 
orbits are also forbidden in this space-time. 

\subsubsection{Radial motion}

The radial motion for photons is characterized by 
the vanishing angular momentum, $L=0$. Thus,
the effective potential (\ref{n1}) is
$ V_{nr}=0$, and therefore only photons with radial
motion can escape to spatial infinity.
Thus, the radial Eq. (\ref{g9}) reduces
to

\begin{equation}\label{n3}
\dot{r}^{2}=\frac{E}{r^{2}},
\end{equation}%
so, an elemental integration leads to
\begin{equation}\label{n4}
\tau \left( r\right) =\pm \frac{R_{0}^{2}}{2\sqrt{E}}\,\left[ \left(
\frac{r}{R_{0}}\right) ^{2}-1\right],
\end{equation}%
where $R_{0}$ denotes the radial distance 
of the massless particle when $\tau=0$. 
This result is in accordance with previous works
dealing with other Lifshitz space-times:
\cite{Cruz:2013ufa,vv13,Maeda:2011jj}.

\begin{figure}[!h]
 \begin{center}
  \includegraphics[width=85mm]{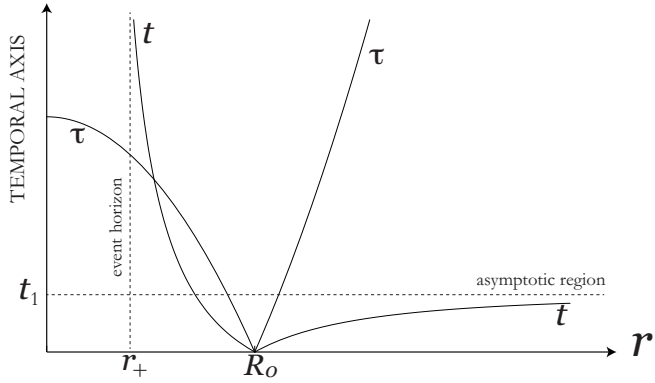}
 \end{center}
 \caption{Plot of the radial motion of massless particles.
 Particles moving to event horizon, $r_+$, cross in a finite
 proper time, but an external observer will see that photons
 take an infinite (coordinate) time to do it.
 Here the values $r_+=2$, $R_0=5$,
 and $E=10^{4}$ have been employed (in geometrized units).}
 \label{f2}
\end{figure}

Now, integrating Eq.  (\ref{g10}) 
an explicit expression for the coordinate time is easily obtained:

\begin{equation}\label{n5}
t\left( r\right) =\pm \frac{1}{4r_{+}^{2}}\ln \left( \frac{\left(
r^{2}-r_{+}^{2}\right) \left( R_{0}^{2}+r_{+}^{2}\right) }{\left(
r^{2}+r_{+}^{2}\right) \left( R_{0}^{2}-r_{+}^{2}\right) }\right) .
\end{equation}%
In the asymptotic region, $r\rightarrow \infty$, we obtain the
limit
\begin{equation}\label{n6}
t_{1}=\lim_{r\rightarrow \infty }t\left( r\right) =\frac{1}{4r_{+}^{2}}\ln
\left( \frac{R_{0}^{2}+r_{+}^{2}}{R_{0}^{2}-r_{+}^{2}}\right) .
\end{equation}%
This fact has been reported in other Lifshitz black holes
\citep{Olivares:2013uha,Cruz:2013ufa}, and seems to express  a behavior
characteristic of this kind of space-time (\citet{vv13}).
Fig. \ref{f2} shows this situation graphically.

\subsubsection{Angular motion}

Massless particles with non-vanished angular momentum, ($L\neq 0$),
comply with the effective potential given by

\begin{equation}\label{n7}
V_{na}\left( r\right) =L^{2}\left( \frac{r^{4}-r_{+}^{4}}{r^{2}}\right);
\end{equation}%
thus, the turning point is located at%
\begin{equation}
R_{L}=\sqrt{\frac{E}{2L^{2}}+\sqrt{\frac{E^{2}}{4L^{4}}+r_{+}^{4}}}.
\label{n8}
\end{equation}%
Now, using Eq. (\ref{g11}), the quadrature is obtained
\begin{equation}\label{n9}
\theta \left( r\right) =-\int_{R_{L}}^{r}\frac{1}{\sqrt{\left( R_{L}-r\right)
\left( r+R_{L}\right) \left( r+i\rho \right) \left( r-i\rho \right) }}\,dr,
\end{equation}%
where the roots of the fourth-degree polynomial inside the radical is given
by

\begin{equation}\label{n10}
\rho =\sqrt{-\frac{E}{2L^{2}}+\sqrt{\frac{E^{2}}{4L^{4}}+r_{+}^{4}}}.
\end{equation}%
In order to integrate Eq. (\ref{n9}), $u=R_{L}-r$ is set, and
after a brief manipulation, we obtain the polar
trajectory of massless particles,
\begin{equation}\label{n11}
r\left( \theta \right) =R_{L}-\frac{1}{4\wp \left( \sqrt{u_{1}u_{2}u_{3}}\,\theta
;g_{2},g_{3}\right) +\frac{\alpha }{3}},
\end{equation}%
where $\wp \equiv \wp(y; g_2, g_3)$ is the $\wp$-Weierstra{\ss} function, and
$g_{2}$ and $g_{3}$ are the so-called
Weierstra{\ss} invariants given by
\begin{equation}\label{n12}
g_{2}=\frac{1}{4}\left( \frac{\alpha ^{2}}{3}-\beta \right) ,\,g_{3}=%
\frac{1}{16}\left( \gamma +\frac{2}{27}\alpha ^{3}-\frac{\alpha \beta }{3}%
\right).
\end{equation}
The other constants are
\begin{equation}
\alpha =\frac{1}{u_{1}}+\frac{1}{u_{2}}+\frac{1}{u_{3}},\quad
\beta =\frac{1}{u_{1}u_{2}}+\frac{1}{u_{1}u_{3}}+\frac{1}{u_{2}u_{3}},\quad
\gamma =\frac{1}{u_{1}u_{2}u_{3}}.
\end{equation}%
with,
\begin{equation}
u_{1}=2R_{L},\,u_{2}=R_{L}+i\rho ,\,u_{3}=R_{L}-i\rho.
\end{equation}%
\bigskip
\begin{figure}[!h]
 \begin{center}
  \includegraphics[width=85mm]{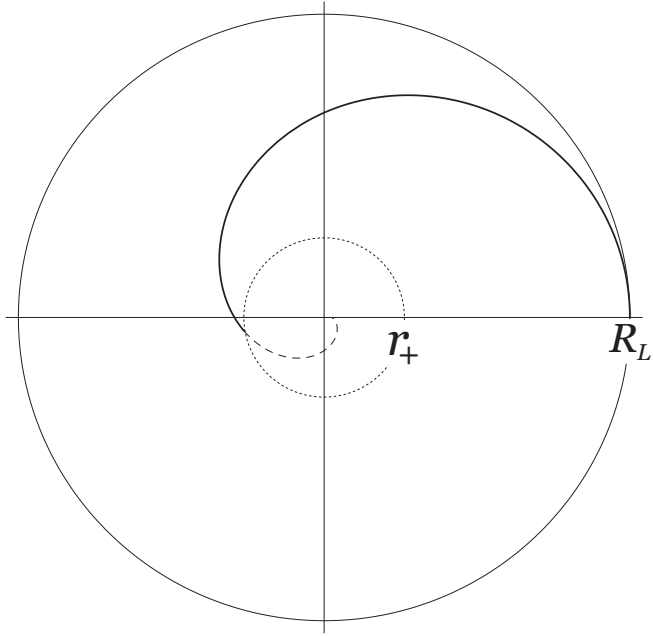}
 \end{center}
 \caption{Plot of the angular motion of massless particles.
 The solid line represents the regular orbit performed
 by the photons, while the dashed line is the analytic 
 continuation of that orbit. 
 Here,  the initial condition 
 $\theta=0$ when $r=R_L$, together with the values
 $L=1$, $r_+=2$, $R_L=8$ (in geometrized
 units), has been employed.}
 \label{f3}
\end{figure}

\subsection{Time-Like Geodesics}

\begin{figure}[!h]
 \begin{center}
  \includegraphics[width=85mm]{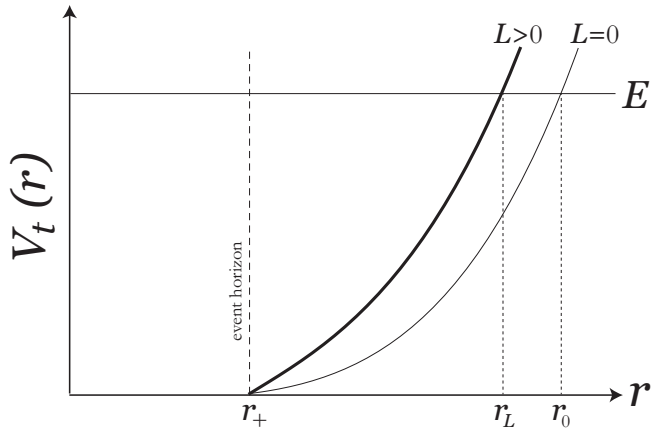}
 \end{center}
 \caption{Plot of the effective potential of massive particles.
 This graph shows that, independently of the angular momentum,
 particles cannot escape to spatial infinity.
 Here  the values $r_+=2$, $L=0$ for radial motion, 
 and $L=4$ for polar motion (in geometrized units) were employed.}
 \label{f4}
\end{figure}

In this section the motion
of massive particles, $m=1$, is computed, so the
effective potential is given by
\begin{equation}
V_{t}\left( r\right) =\left( r^{4}-r_{+}^{4}\right)
\left( 1+\frac{L^{2}}{r^{2}}\right).  \label{t1}
\end{equation}%
This is illustrated in Fig. \ref{f4} for radial ($L=0$) and non-radial ($L\neq0$) particles.

\subsubsection{Radial motion}

\begin{figure}[!h]
 \begin{center}
  \includegraphics[width=85mm]{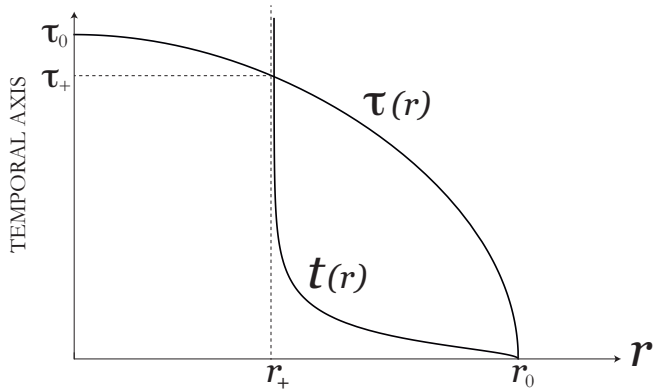}
 \end{center}
 \caption{Plot of the radial motion of massive particles.
 Particles moving to event horizon, $r_+$, cross into a finite
 proper time, but an external observer will see that particles
 take an infinite (coordinate) time to do so.
 Here, $r_+=2$, $r_0=5$ and $E=609$ (in geometrized units)
 have been used.}
 \label{f5}
\end{figure}

In this case the effective potential (\ref{t1}) becomes

\begin{equation}\label{t2}
V_{t}\left( r\right) =\left( r^{4}-r_{+}^{4}\right) ;
\end{equation}%
thus, the turning point is located at
\begin{equation}\label{t3}
r_{0}=\left( E+r_{+}^{4}\right)^{1/4}.
\end{equation}%
Therefore, using Eq. (\ref{t2}) and Eq. (\ref{g9}), the proper time
of the radial massive particles in terms of the radial coordinate $r$
can be obtained explicitly, which results in

\begin{equation}\label{t4}
\tau \left( r\right) =\frac{1}{2}\arccos\left(\frac{r^2}{r_0^2}\right).
\end{equation}%
It is interesting to note that massive particles take a finite proper
time, $\tau_+ \equiv \tau(r=r_+)$, to cross the
event horizon, which depends on the initial distance, $r_0$. Also,
the analytic continuation of the motion
in the region $r<r_+ $ means that it takes a finite proper time
$\tau_0 \equiv \tau(r=0) = \pi/4$ to reach
the singularity, which turns out to be independent of the initial distance $r_0$.
This is a novel result because it does not agree 
with those obtained in previous works 
\citep{Olivares:2013uha,Cruz:2013ufa},  
where the time to the singularity depends 
on the initial distance and the parameters 
of space-time through the event horizon (see Fig. \ref{f5}).

On the other hand, Eq. (\ref{t2}) into Eq. (\ref{g10})
yields the coordinate time directly as a function of $r$,

\begin{equation}\label{t5}
t\left( r\right) =\frac{1}{4r_{+}^{2}}
\left[ \textrm{arccosh}\left(\frac{r_0^4-r_+^2\, r^2}{r_0^2\,(r^2-r_+^2)}\right)
-\textrm{arccosh}\left(\frac{r_0^4+r_+^2 \,r^2}{r_0^2\,(r^2+r_+^2)}\right)\right].
\end{equation}
In Fig. \ref{f5} we plot the functional relations (\ref{t4}) and
(\ref{t5}), which show us that the physics is essentially the same as 
Einstein's space-times (S, SdS, SAdS, etc.).

\subsubsection{Angular motion}

\begin{figure}[!h]
 \begin{center}
  \includegraphics[width=85mm]{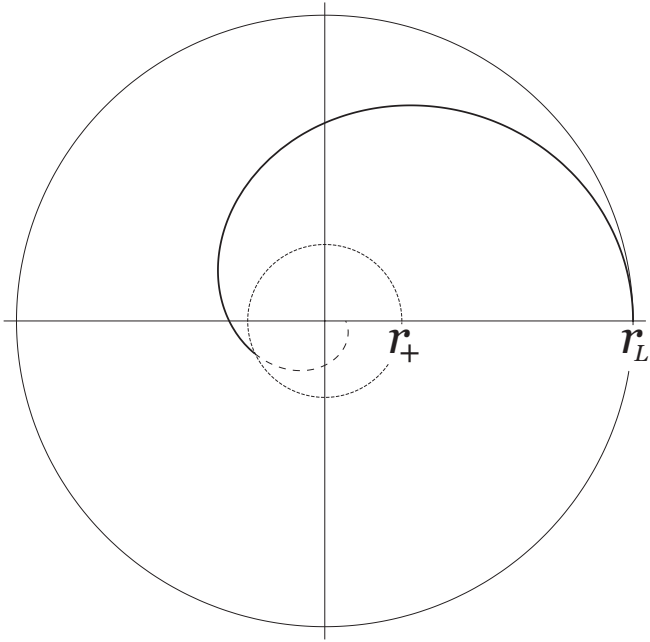}
 \end{center}
 \caption{Polar trajectory for massive particles.
 The solid line represents the regular orbit performed by the test particles, while the dashed line is the analytic continuation of that orbit. Here the initial condition $\theta=0$, when $r=r_L$,
 together with the values $L=1$, $r_+=2$, $r_L=8$ and $E=4\times10^{3}$ 
 ( in geometrized units) has been employed.}
 \label{f6}
\end{figure}

In the case of massive particles with non-vanished angular momentum, we have
that effective potential is given by

\begin{equation}\label{t6}
V_{ta}\left( r\right) =\left( r^{4}-r_{+}^{4}\right)
\left( 1+\frac{L^{2}}{r^{2}}\right);
\end{equation}%
thus, we can write Eq. (\ref{g11}) as
\begin{equation}\label{t7}
  \left(\frac{dr}{d\theta}\right)^2=\frac{(r_L^2-r^2)(r^2-r_1^2)(r^2-r_2^2)}{L^2}.
\end{equation}
Here $r_L$ is the turning point given by the relation
\begin{equation}\label{t8}
  r_L^2=u_0-\frac{L^2}{3},
\end{equation}
and $r_1$, $r_2$ are two complex quantities (without physical meaning)
given by
\begin{equation}\label{t9}
  r_j^2=u_j-\frac{L^2}{3},\qquad (j=1, 2)
\end{equation}
with
\begin{equation}\label{t10}
  u_n=\sqrt{\frac{\eta_2}{3}}\cos\left[\frac{1}{3}
  \arccos\sqrt{\frac{27\,\eta_3^2}{\eta_2^3}}+\frac{2\,n\,\pi}{3}\right],\qquad (n=0, 1, 2)
\end{equation}
where the $\eta$'s are given by
\begin{eqnarray}\label{t11}
  \eta_2 &=& 4\left(E+r_+^4+\frac{L^4}{3}\right)\quad (>0) \\ \label{t12}
  \eta_3 &=&  -4\left(\frac{2 L^6}{27}+\frac{L^2}{3}(E+r_+^4)-L^2 r_+^4\right)\quad (<0).
\end{eqnarray}
Therefore, after a little algebraic manipulation in Eq. (\ref{t7}),
the polar trajectory of the massive particles in terms of the $\wp$-Weierstra{\ss} function
is obtained, which results in

\begin{equation}\label{t13}
r\left( \theta \right) =\sqrt{r_{L}^{2}-
\frac{1}{4\wp \left(\frac{2}{L}\sqrt{y_{1}\,y_{2}\,y_{3}}\,\theta ;g_{2}, g_{3}\right) +\frac{\overline{\alpha} }{3}}},
\end{equation}
where the Weierstra{\ss} invariants are given by
\begin{equation}\label{t14}
g_{2}=\frac{1}{4}\left( \frac{\overline{\alpha} ^{2}}{3}-\overline{\beta} \right) ,\,g_{3}=%
\frac{1}{16}\left( \overline{\gamma} +\frac{2}{27}\overline{\alpha} ^{3}-\frac{\overline{\alpha} \overline{\beta} }{3}%
\right),
\end{equation}%
with
\begin{equation}\label{t15}
\overline{\alpha}=\frac{1}{y_1}+\frac{1}{y_2}+\frac{1}{y_3} ,
\qquad\overline{\beta} =\frac{1}{y_{1}\,y_{2}}+\frac{1}{y_{1}\,y_{3}}+\frac{1}{y_{2}\,y_{3}},\,
\overline{\gamma} =\frac{1}{y_{1}\,y_{2}\,y_{3}}.
\end{equation}%
Also,  we have that
\begin{equation}\label{t16}
  y_1=r_L^2,\quad y_2=r_L^2-r_1^2,\quad \textrm{and}\quad y_3=r_L^2-r_2^2.
\end{equation}
In Fig. \ref{f6} we plot the polar trajectory (\ref{t13}).

\section{Final Remarks}

In this paper the geodesic structure of a Lifshitz 
black hole that is a solution to the Einstein-scalar-Maxwell 
theory in $3+1$ space-time dimensions with 
critical exponent  $z=2$ was examined. 
Using the Lagrangian procedure radial and 
angular motions of massless and massive 
test particles were studied. Analytical expressions 
for the proper time and coordinate time as a function 
of the radial coordinate for the strictly radial motion 
were obtained and the same analysis was provided for the 
polar trajectories.
In Fig. \ref{f2}, the proper and coordinate time for 
radial photons was depicted. 
The graphic shows that an external observer can see
that it takes a finite coordinate time, $t_1$, for the 
photons to reach the asymptotic region, but 
an infinite proper time to do it. This result is consistent 
with previous studies dealing with Lifshitz 
space-times: a topological black hole of $3+1$ 
dimensions with critical exponent  $z = 2$
\citep{Olivares:2013uha}, in $2+1$ dimensions and $z=3$ 
\citep{Cruz:2013ufa}, generalized for $D$-dimensions, and
with an arbitrary critical exponent $z$ \citep{vv13}.
It was also found that the massless and massive particles (see Fig. \ref{f5}) 
cross the event horizon, $r_+$, in a finite proper time.
However, the external observer can see that it takes the photon an infinite
time to reach the horizon, which is analogous to the situation that occurs with the Einstein
space-times. Angular motions are described by the polar
trajectories of the particles in terms of the $\wp$-Weierstra{\ss} function, the results of which
are depicted in Fig. \ref{f3} for photons and in Fig. \ref{f6} for massive particles.
The behavior is analogous for both types of particles,
where the motion is studied on the invariant plane $\phi= \pi/2$, and it is observed that bounded orbits are not allowed for this space-time on this plane. Similar behavior has 
recently been reported in the literature for other asymptotically 
Lifshitz black holes \citep{Olivares:2013uha,Cruz:2013ufa}.

\begin{acknowledgements}
The authors would like to thank F. Ercol\'i, M. Cur\'e and  referees for their comments 
and suggestions that helped us improve this paper.
This work was funded by the Comisi{\'o}n Nacional de Investigaci{\'o}n Cient{\'i}fica y Tecnol{\'o}gica through FONDECYT Grants 11121148 (YV, FM) and 11130695 (JRV), and also partially funded by the Direcci{\'o}n de Investigaci{\'o}n, Universidad de La Frontera (FM). 
MO would also like to thank PUCV and Instituto de F\'isica y Astronom\'ia (UV) and
JRV wishes to thank to the UFRO for their hospitality.
\end{acknowledgements}



\begin{thebibliography}{99}

\bibitem[\protect\citeauthoryear{Amado and Faedo}{2011}]{Amado:2011nd} 
Amado I., Faedo A. F.:
Lifshitz black holes in string theory.
  JHEP {\bf 1107}, 004 (2011)
  [arXiv: 1105.4862].

\bibitem[\protect\citeauthoryear{Ayon-Beato et al.}{2009}]{AyonBeato:2009nh}
Ayon-Beato E.,~Garbarz A.,~Giribet G. and
~Hassaine M.: Lifshitz Black Hole in Three Dimensions,
Phys.\ Rev.\ D \textbf{80}, 104029 (2009).
  
\bibitem[\protect\citeauthoryear{Balasubramanian and McGreevy}{2009}]{Balasubramanian:2009rx}
Balasubramanian K., McGreevy J.: An
Analytic Lifshitz black hole. Phys.\ Rev.\ D \textbf{80}, 104039 (2009).


\bibitem[\protect\citeauthoryear{Brynjolfsson et al.}{2010}]{Brynjolfsson:2009ct} 
  Brynjolfsson E. J., Danielsson U.~H., Thorlacius L., Zingg T.:
Holographic Superconductors with Lifshitz Scaling.
  J.\ Phys.\ A {\bf 43}, 065401 (2010)
  [arXiv:0908.2611].
  

\bibitem[\protect\citeauthoryear{Cruz et al.}{2013}]{Cruz:2013ufa}
Cruz N., Olivares M. and Villanueva J. R.: Geodesic
Structure of Lifshitz Black Holes in 2+1 Dimensions.
Eur. Phys. J. C  {\bf 73}, 2485 (2013) [arXiv: 1305.2133].

\bibitem[\protect\citeauthoryear{Cruz et al.}{2005}]{Olivares} 
Cruz N., Olivares M. and Villanueva J.R.:
The Geodesic Structure of the Schwarzschild anti-de Sitter Black Hole.
Class. Quantum Grav. \textbf{22}, 1167-1190 (2005).
 [arXiv: 0408016]


\bibitem[\protect\citeauthoryear{Dehgani and Mann}{2010}]{Dehghani:2010kd}
Dehghani M.~H. and~Mann R.~B.: Lovelock-Lifshitz
Black Holes. JHEP \textbf{1007}, 019 (2010).

\bibitem[\protect\citeauthoryear{Devecioglu and Sarioglu}{2011}]{Devecioglu:2011yi} 
Devecioglu D. O. and. Sarioglu O.:
On the thermodynamics of Lifshitz black holes.
Phys.\ Rev.\ D {\bf 83}, 124041 (2011)
[arXiv: 1103.1993 ].






\bibitem[\protect\citeauthoryear{Hartnoll et al.}{2010}]{Hartnoll:2009ns}
Hartnoll S. A., Polchinski J., Silverstein E.,
Tong D.:  Towards strange metallic
holography.\ JHEP \textbf{1004}, 120 (2010).




\bibitem[\protect\citeauthoryear{Kachru et al.}{2008}]{Kachru:2008yh}
Kachru S., Liu X., Mulligan M.:
Gravity Duals of Lifshitz-like Fixed Points,\ Phys.\ Rev.\
D \textbf{78}, 106005 (2008).


  

\bibitem[\protect\citeauthoryear{Maeda and Giribet}{2013}]{Maeda:2011jj}
Maeda H. and Giribet G.:
Lifshitz black holes in Brans-Dicke theory.
JHEP {\bf 1111}, 015 (2011) 
[arXiv: 1105.1331].

\bibitem[\protect\citeauthoryear{Maldacena}{1998}]{Maldacena:1997re}
Maldacena J.~M.:  The Large N
limit of superconformal field theories and supergravity.
Adv.\ Theor.\ Math.\ Phys.\ \textbf{2}, 231 (1998).


\bibitem[\protect\citeauthoryear{Mann}{2009}]{mann} 
Mann R. B.: 
Lifshitz topological black holes. 
JHEP \textbf{06}, 075 (2009).

\bibitem[\protect\citeauthoryear{Myung and Moon}{2012}]{Myung:2012cb} 
Myung Y. S. and Moon T.:
Quasinormal frequencies and thermodynamic quantities for the Lifshitz black holes.
Phys.\ Rev.\ D \textbf{86}, 024006 (2012)
[arXiv: 1204.2116].

\bibitem[\protect\citeauthoryear{Myung}{2012}]{oai:arXiv.org:1203.1367} 
 Myung Y. S.:
Phase transitions for the Lifshitz black holes.
  Eur.\ Phys.\ J.\ C {\bf 72}, 2116 (2012)
  [arXiv: 1203.1367].




\bibitem[\protect\citeauthoryear{Olivares et al.}{2013}]{Olivares:2013uha}
Olivares M., Rojas G., V\'{a}squez Y. and Villanueva J. R.: 
Particles motion on topological Lifshitz black holes in 3+1 dimensions. 
Astrophys. Space Sci. \textbf{347}, 83-89 (2013) 
[arXiv: 1304.4297].

\bibitem[\protect\citeauthoryear{Olivares and Villanueva}{2013}]{ov13}
Olivares M. and Villanueva J. R.:
Massive neutral particles on heterotic string theory.
Eur. Phys. J. C \textbf{73},  2659 (2013)
[arXiv: 1311.4236]

\bibitem[\protect\citeauthoryear{Pang}{2010}]{Pang:2009wa}
Pang D.~W.: Conductivity and Diffusion Constant in
Lifshitz Backgrounds. JHEP \textbf{1001}, 120 (2010).







\bibitem[\protect\citeauthoryear{Taylor}{2008}]{Taylor:2008tg}
Taylor M.: Non-relativistic holography.
(2008) [arXiv: 0812.0530].








\bibitem[\protect\citeauthoryear{Villanueva and Olivares}{2013}]{villanueva1}
Villanueva J. R. and Olivares M.:
On the null trajectories in conformal Weyl Gravity.
J. Cosmol. Astropart. Phys.
{\bf 1306}, 040 (2013)
[arXiv: 1305.3922].



\bibitem[\protect\citeauthoryear{Villanueva and V\'asquez}{2013}]{vv13}
Villanueva J. R. and V\'asquez Y.:
About the coordinate time in Lifshitz space-times.
Eur. Phys. J. C \textbf{73},  2587 (2013)
[arXiv: 1309.4417].

\end{thebibliography}
\end{document}